\documentclass{svproc}
\usepackage{graphicx}%
\usepackage{multirow}%
\usepackage{amsmath,amssymb,amsfonts}%
\usepackage{mathrsfs}%
\usepackage[title]{appendix}%
\usepackage{xcolor}%
\usepackage{textcomp}%
\usepackage{manyfoot}%
\usepackage{booktabs}%
\usepackage{algorithm}%
\usepackage{algorithmicx}%
\usepackage{algpseudocode}%
\usepackage{listings}%
\usepackage{url}

\begin{document}
\mainmatter              
\title{Examining I2P Resilience: Effect of Centrality-based Attack}
\titlerunning{I2P Resilience}  
%
\author{Kemi Akanbi\and Sunkanmi Oluwadare\and
Jess Kropczynski \and Jacques  Bou Abdo}
\authorrunning{Kemi Akanbi et al.} 

\institute{Institution1, address,\\
\email{akanbikr@mail.uc.edu.email},\\ }
\institute{University of Cincinnati, OH, United States\\}

\maketitle              

\begin{abstract}

This study examines the robustness of I2P, a well-regarded anonymous and decentralized peer-to-peer network designed to ensure anonymity, confidentiality, and circumvention of censorship. Unlike its more widely researched counterpart, TOR, I2P’s resilience has received less scholarly attention. Employing network analysis, this research evaluates I2P’s susceptibility to adversarial percolation. By utilizing the degree centrality as a measure of nodes' influence in the network, the finding suggests the network is vulnerable to targeted disruptions. Before percolation, the network exhibited a density of 0.01065443 and an average path length of 6.842194. At the end of the percolation process, the density decreased by approximately 10\%, and the average path length increased by 33\%, indicating a decline in efficiency and connectivity. These results highlight that even decentralized networks, such as I2P, exhibit structural fragility under targeted attacks, emphasizing the need for improved design strategies to enhance resilience against adversarial disruptions.
\end{abstract}

\section{Introduction}\label{sec1}

The increasing need of individuals to remain anonymous and ensure the confidentiality of shared information has spurred the development of various privacy-enhancing technologies, including I2P, TOR, and ZeroNET \cite{Hoang2018}. The Invisible Internet Project (I2P) is a decentralized anonymous overlay network with low latency \cite{JP2012}, established in 2003 to safeguard anonymity, ensure confidentiality, and bypass censorship \cite{Roberto_Magan-Carrion}. To bolster its anonymity and confidentiality features, I2P employs garlic routing, an enhanced version of TOR's onion routing, where messages are bundled into a "garlic bulb" \cite{JP2012}. To participate in the network, peers register their routers as nodes, contributing their bandwidth for network-wide communication \cite{Roberto_Magan-Carrion}.

I2P has been examined through multiple lenses. For instance, Wilson and Bazli undertook a “forensic analysis of I2P activities” \cite{wilson2016forensic}. Empirical studies have been conducted to describe network properties, such as churn rate, bandwidth distribution, geographical distribution, resilience to attacks, and censorship \cite{PeipingLiu,LinkenLiu2019,Hoang2018}. Bou Abdo and Hussain \cite{Abdo/978-3-031-53472-0_30} mathematically modeled the network to illustrate its degree distribution and validated their model with agent-based simulation. Given its dynamic nature, understanding the network's evolving properties is crucial. Moreover, the network has experienced substantial growth, expanding from 25,795 nodes as of January 2019 to 72,653 nodes in April 2025, as indicated on the I2P metrics page \cite{i2pmetrics}. This expansion could introduce potential risks of centralization within the decentralized system. Consequently, this study focuses on analyzing the structure and resilience of I2P against targeted removal attacks using network analysis techniques.

Network science has emerged as an indispensable methodology, especially in the age of big data generated across diverse domains \cite{Evans2022}. The real strength of network analysis is in its ability to reveal emergent behaviors stemming from interaction patterns at meso and macro scales, extending beyond mere individual connections \cite{Evans2022}. To gain insight into the I2P resilience, this study measures the degree of centrality, identifies key players, and subsequently simulates the removal of the three key players to address the research question: How susceptible is I2P to targeted attacks aimed at eliminating high-centrality nodes? 

The remaining sections of this paper are structured as follows: Section 2 is the Literature Review, which provides an examination of prior research and methodologies applied in the study of the I2P. Section 3 describes the Methodology, detailing the data acquisition, cleaning, and analysis procedures. Section 4 presents Results and Discussion. Lastly, Section 5 is the Conclusion, which summarizes the key insights of the research and proposes a future direction.

\section{Literature Review}\label{sec2}

Since the development of I2P, several studies have attempted to understand the network. Bou Abdo \cite{Abdo/978-3-031-53472-0_30} introduced a mathematical model for I2P's peer selection mechanism. The study describes the probability of weighted node selection and recursively models the resulting network's degree distribution, allowing for predictions about network structures and attributes based on its parameters. An agent-based simulation was used to validate the mathematical model, thereby establishing a theoretical basis for analyzing I2P's resilience and peer selection process \cite{Abdo/978-3-031-53472-0_30}.

To better understand the network, several studies have conducted empirical analyses of the I2P network. Authors in \cite{Hoang2018} conducted an empirical measurement of the I2P anonymity network over a three-month period, utilizing 20 routers to collect data on network properties, including the number of active peers and their geographical distribution. The findings demonstrated that, despite I2P's decentralized architecture, it is highly vulnerable to censorship, highlighting the ease with which access can be restricted and the implications for the network's resilience \cite{Hoang2018}.

Authors in \cite{Hu} analyzed the effectiveness of Machine Learning and Deep Learning models in predicting the type of traffic and user behavior in I2P. 
While significant research efforts have been dedicated to analyzing I2P's traffic and services, studies investigating the structural properties and resilience of the network through the lens of network science are limited. 

The study \cite{Siddique_resilience} modeled I2P as a scale-free network and found that the network collapsed under targeted attacks after only 25–35\% of key nodes were removed. At the same time, I2P Prime, a proposed topology, remained intact and resistant to fragmentation until nearly 50\% of its nodes were lost. This shows that I2P resilience is closely tied to its network structure. However, there are limited studies on I2P at the network layer \cite{PeipingLiu}.

Several studies \cite{Wang,Hoang2018}, including a recent work \cite{Siddique}, identify the United States, Russia, and Germany as the dominant hubs. Hence, this study will examine the resilience of the I2P network to centrality-based attacks from a social network analysis perspective, thereby contributing to the knowledge about the network.

\section{Methodology}\label{sec3}

The method adopted in this study is social network analysis. The section describes the data acquisition, preprocessing, and analysis tools, as well as the evaluation metrics for peer influence on the network. 

The study utilizes a subset of data collected on the live I2P network in our empirical work. To clean the dataset, the routers without their IP address displayed were removed, and consequently, the nodes connected to them in the tunnel were also removed from the dataset. In this context, nodes represent other routers within the I2P network, each uniquely identified by an IP address, and participate in tunnels that move traffic from one router to another. 

According to \cite{Roberto_Magan-Carrion}, the I2P network can be mathematically described as a “directed graph G = (V, E),” where V represents the set of routers (or nodes) and E denotes the connections (or edges) between them. Each edge (u, v) indicates a direct link from router u to router v. For our network analysis, we designated the “Scr IP” as the source node and the “Dst IP” as the target node. The dataset utilized consists of 3,081 nodes and 101,105 edges. The following tools were used for the analysis and visualization. “”

 \begin{itemize}
    \item igraph, a “software package for complex network research”\cite{igraph_Csardi_2006}.
    \item RStudio, an "Integrated Development Environment for R"\cite{Rstudio} 
 \end{itemize}

\subsection{ Network structure Analysis}\label{subsec2}

This subsection aims to explore and describe the fundamental structural characteristics of I2P by analyzing the I2P network dataset through the lens of social network analysis, to gain insights that will help answer our research question. Understanding these structural properties is a crucial step in identifying the key players within the network that will be the focus of our targeted node removal attack. 

The formation of triangles occurs in a network, and one common way to measure this is by calculating transitivity. Transitivity measures the percentage of connected pairs of nodes that are also part of a triangle \cite{Seshadhri_Transitivity}. We found a transitivity of  0.003351906. However, our graph is directed, and transitivity alone does not provide a clear picture of the network's structure \cite{Seshadhri_Transitivity}.
 \begin{figure}
     \centering
     \includegraphics[width=0.90\linewidth]{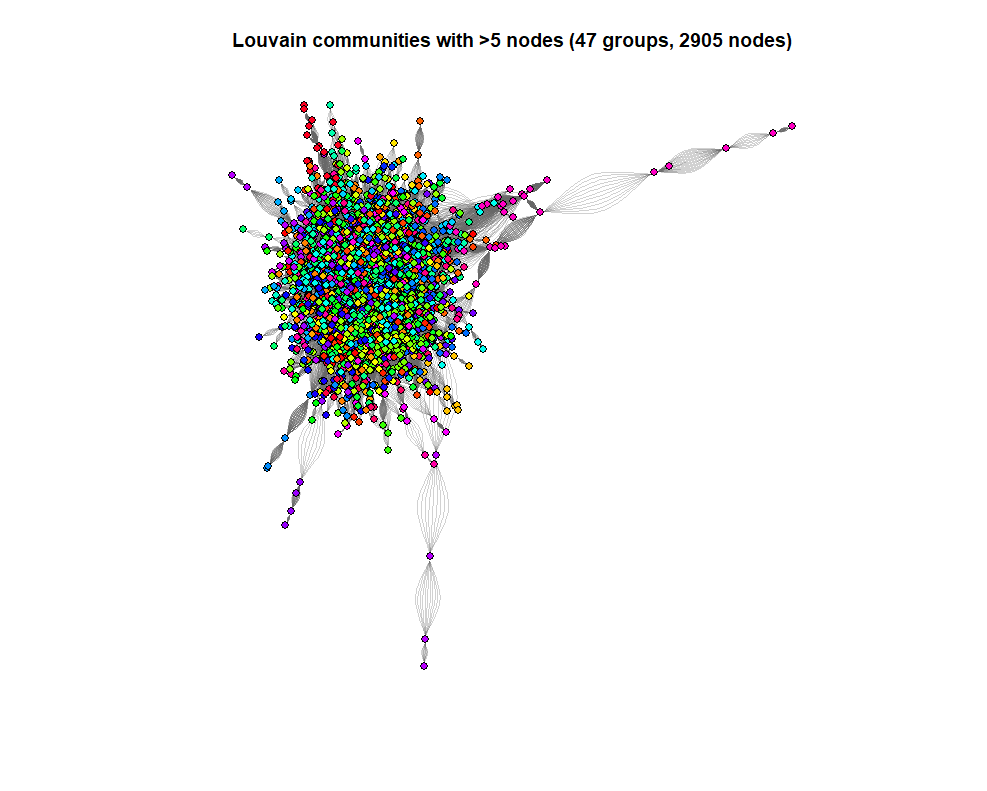}
     \caption{Detected Communities}
     \label{fig:placeholder}
 \end{figure}
 
 Hence, we employed the Louvain community detection algorithm on the network data and discovered 128 groups with a modularity of 0.76. Modularity measures the number of connections inside these groups, which is much higher than what you would expect randomly \cite{Newman}. A positive modularity score indicates that the network is indeed organized into well-defined communities \cite{Newman}. Our analysis yielded a high modularity score of 0.76, confirming that the network exhibits clear community structure. Figure 1 illustrates the groups with more than five members found in the network.

\subsection{Percolation}\label{sec4}

It is essential to discuss the concept of percolation to assess the impact of removing central nodes as a targeted attack on I2P. Percolation is defined by \cite{Rapisardi2022-homeostatic} as a process that disrupts network connectivity by removing nodes or links, providing a valuable framework to understand how the structure of a network contributes to its robustness.

Authors in \cite{PENG201817} describe the degree centrality as the number of edges connected to a particular node in the network. In the case of directed networks, this measure is often split into two components: in-degree, which counts the number of edges directed toward a node, and out-degree, which counts the number of edges a node directs to others. Among the vast array of centrality measures proposed over the years, Degree Centrality, Closeness Centrality, Betweenness Centrality, and Eigenvector Centrality remain central due to their frequent and practical application in understanding node influence and importance in various networked systems \cite{Borgatti2005,PENG201817}. They are considered foundational in network analysis \cite{PENG201817}.  

The distributed architecture of the garlic network is designed to handle random removal of nodes (churn). Complex networks generally exhibit a fascinating ability to withstand both accidental failures and deliberate attacks \cite{Rapisardi2022-homeostatic}. Although failures randomly eliminate network components, attacks strategically target influential nodes or edges to inflict maximum damage \cite{Almeira_centralized_attack}. In attacks targeting network centrality, influential nodes are systematically eliminated based on their importance, quantified by a specific centrality metric, with the most central nodes removed first \cite{Almeira_centralized_attack}. This strategic removal can inflict far greater damage than random disruptions by exploiting the critical roles these nodes play in network connectivity and function. 

For the percolation exercise, we employed both manual and automated techniques. In the manual removal process, we did not recalculate the centrality scores after each removal. Instead, we evaluated the degree of centrality of the top ten nodes central to the network and removed the three most central nodes. On the other hand, the automated approach utilized a for-loop, where the degree centrality is recalculated after removing the most central node. This process ensures that the next node selected for removal in each iteration has the highest degree among all nodes. 
  
\section{Result and Discussion}

Node centrality can be regarded as an effort to measure the structural importance of actors in a network. To analyze and evaluate the influence of a node within a network, employing network measures like centrality measures is a starting point \cite{BORGATTI2006}. Our analysis reveals a network of 3,081 nodes and 101,105 edges with a transitivity measure of 0.003351906, indicating a very low number of possible triangles. However, the Louvain community detection algorithm identified 128 distinct communities within the graph, with a modularity score of 0.76, indicating that the identified groups are tightly connected within themselves and have very few connections to other groups.

Some key metrics are essential for understanding centralization within a network. One such metric is the degree of centralization score. The high score of 0.8194817 observed in this study indicates that the network’s degree distribution is heavily skewed toward a small number of highly connected hubs. This concentration of connectivity makes the network vulnerable to targeted attacks. The modularity score further reinforces this finding by revealing well-defined clusters of nodes within the network. The network’s susceptibility to centrality-based attacks was confirmed by the drop in density and the substantial increase in average path length observed during our percolation exercise.

To pinpoint the most influential actors within our network, we employed degree centrality. By calculating the total degree centrality, we focused on identifying nodes that receive a high number of incoming and outgoing connections, suggesting they are influential within our network. Table 1 presents the in-degree, out-degree centrality, and all-degree centrality of nodes in the network with the last octet of the IP address redacted. The central node, IP address 23.137.254.xxx exhibits the highest in-degree of 2571 and significantly the highest out-degree of 2541 among the top ten nodes, indicating a hub with a large number of incoming connections and almost equal outgoing connections within the network. Interestingly, the IP is located in the United States, consistent with findings from \cite{Wang,Hoang2018,Siddique}, which identifies that the United States is the major contributor to the network. The Degree centrality of the top ten nodes, presented by country, is shown in Figure 2.

 \begin{table}[!ht]
 \caption{Top nodes by Degree}\label{tab1}%
    \centering
    \begin{tabular}{llllll}
        \hline
        IP Address & InDegree & IP Address & OutDegree &  IP Address & TotalDegree \\ \hline
        23.137.254.xxx & 2571 & 23.137.254.xxx & 2541 & 23.137.254.xxx & 5112 \\ 
        185.148.1.xxx & 1514 & 185.148.1.xxx & 1495 & 185.148.1.xxx & 3009 \\ 
        62.60.150.xxx & 1441 & 69.243.247.x & 888 & 62.60.150.xxx & 2237 \\ 
        69.243.247.x & 872 & 62.60.150.xxx & 796 & 69.243.247.x & 1760 \\ 
        78.139.70.xx & 839 & 78.139.70.xx & 650 & 78.139.70.xx & 1489 \\ 
        83.31.208.xxx & 538 & 45.82.122.xxx & 640 & 45.82.122.xxx & 901 \\ 
        84.251.40.xx & 477 & 78.107.238.xx & 618 & 83.31.208.xxx & 698 \\ 
        185.252.177.xx & 448 & 91.198.115.xx & 555 & 185.252.177.xx & 694 \\ 
        103.167.234.xx & 433 & 162.218.65.xx & 553 & 78.107.238.xx & 687 \\ 
        85.130.157.xxx & 384 & 82.130.24.xxx & 419 & 91.198.115.xx & 653 \\ \hline
    \end{tabular}
\end{table}

\begin{figure}
    \centering
    \includegraphics[width=0.75\linewidth]{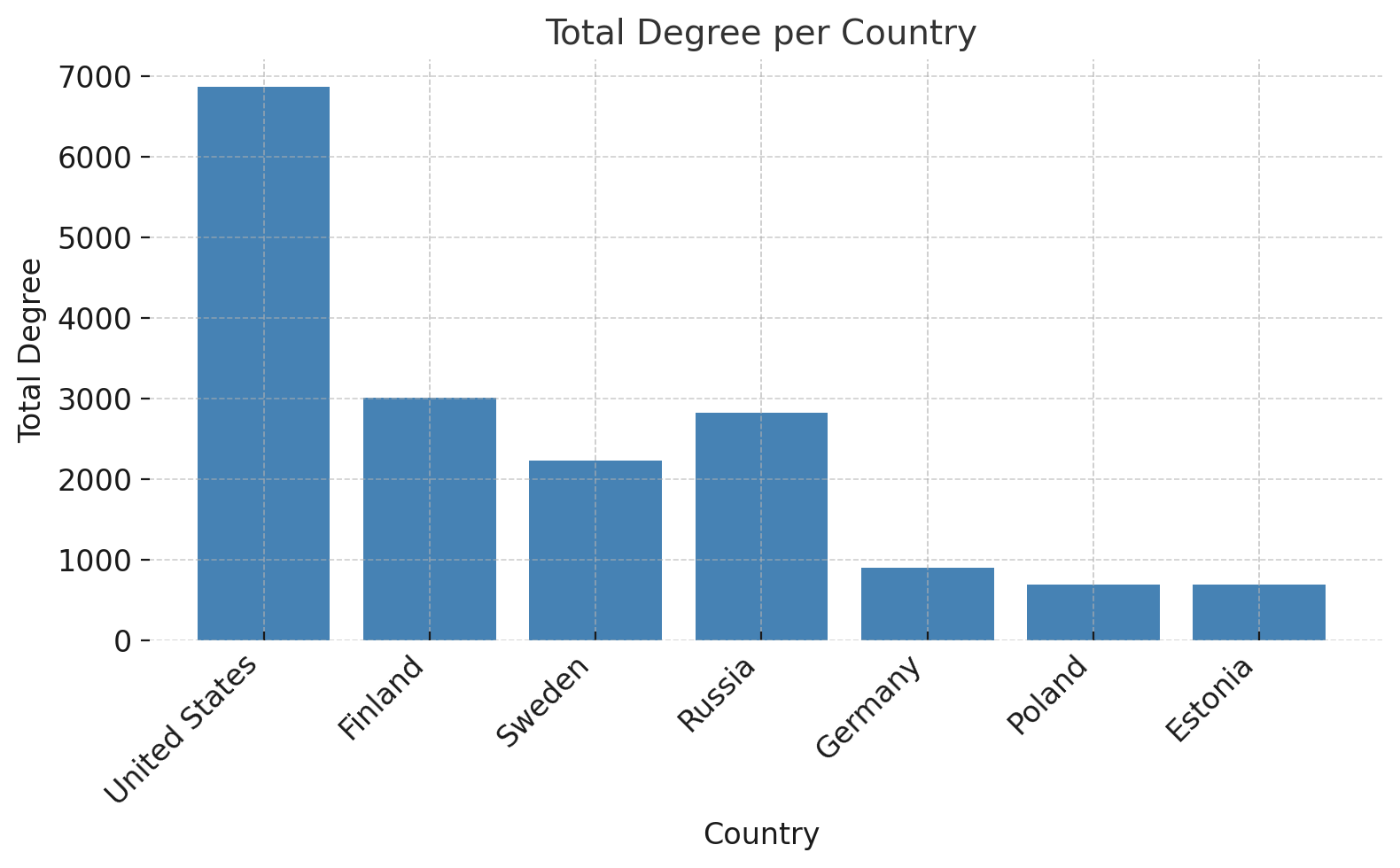}
    \caption{Degree per Country for Top 10 Central Nodes}
    \label{fig:placeholder}
\end{figure}

\begin{figure}[!ht]
    \centering
    \begin{minipage}{0.48\linewidth}
        \centering
        \includegraphics[width=\linewidth]{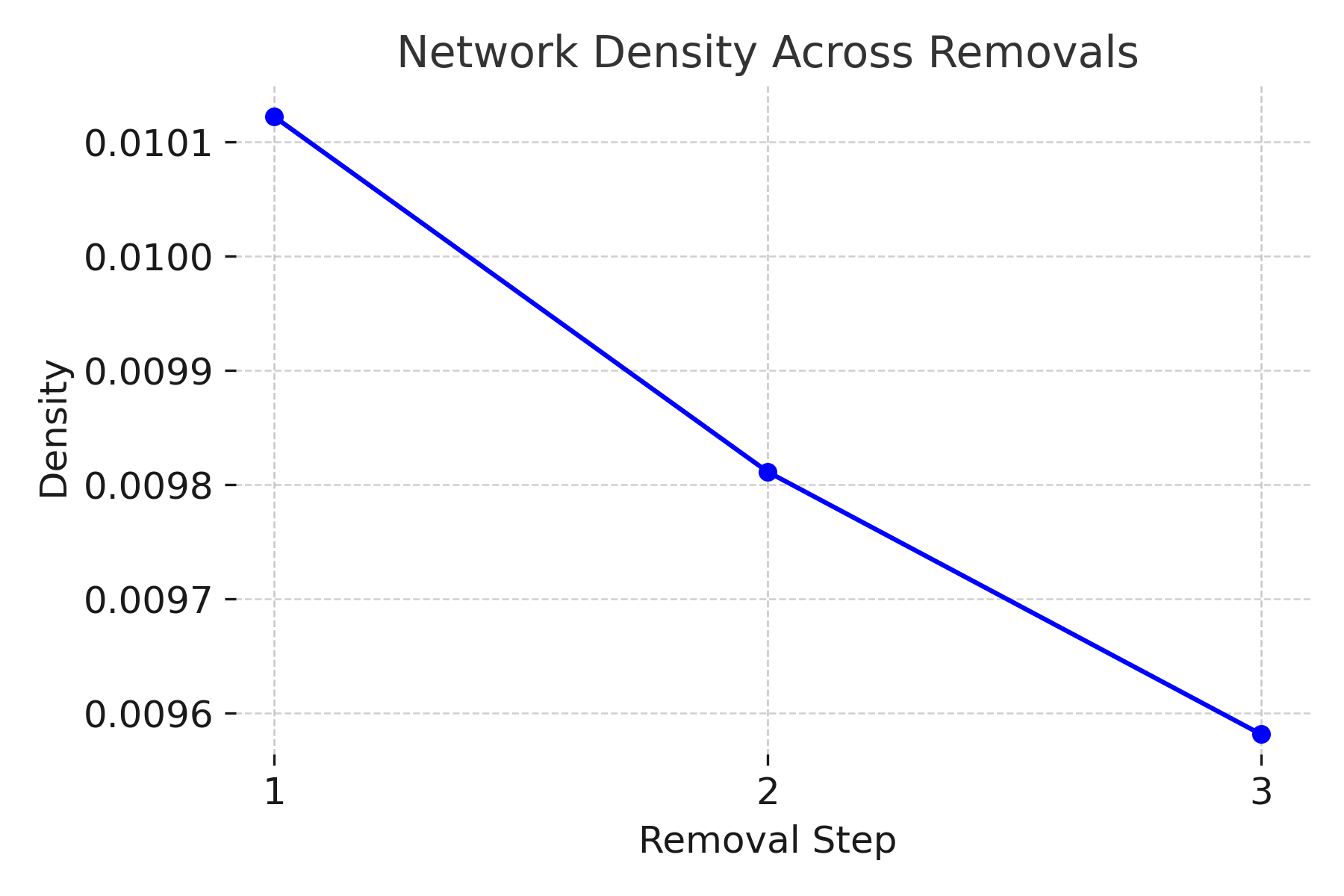}
        \caption{Network Density over Removal}
        \label{fig:density_removal}
    \end{minipage}
    \hfill
    \begin{minipage}{0.48\linewidth}
        \centering
        \includegraphics[width=\linewidth]{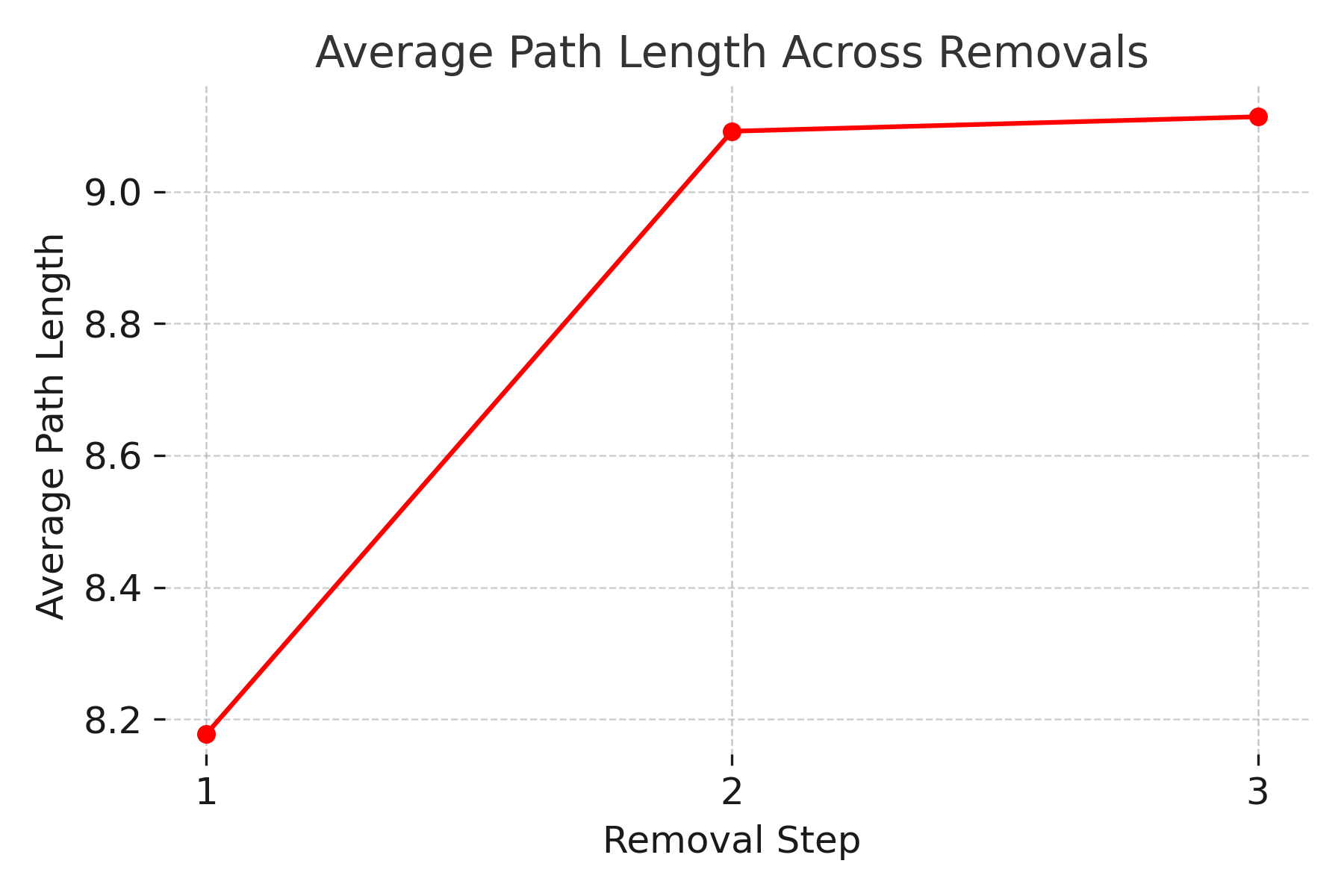}
        \caption{Avg. Path Length Over Removal}
        \label{fig:apl_removal}
    \end{minipage}
\end{figure}

Before the percolation analysis, the network metrics recorded were a density of 0.01065443 and an average path length of 6.842194. Both manual and automated processes identified the same set of nodes as the top three most central. After the removal of the first node, 23.137.254.xxx, the density decreased to 0.0101223, while the average path length increased drastically to 8.17743. With the removal of the second node, 185.148.1.xxx, we observed a significant decline in density to 0.009811376, while the average path length increased to 9.091371. Following the removal of the third node, 62.60.150.xxx, the density decreased to 0.009581559, and the average path length increased to 9.113329. 

The sequential removal of these three key players had a profound impact on the network's structural properties.  The changes in network density following the targeted removal are illustrated in Figure 3, showing a decrease in network density from 0.01065 to 0.00958, indicating a steady loss of connectivity and an increasingly sparse network. Similarly, Figure 4 shows that the average path length increased from 6.84 to 9.11, suggesting that more steps are required to communicate effectively. This combination of decreasing density and increasing path lengths reflects a reduction in network cohesion and efficiency, characteristic of declining robustness. 

The vulnerability of the network becomes even more concerning in the presence of a global passive adversary capable of monitoring large portions of the network over extended periods. Such an adversary can perform traffic analysis, gradually map out the topology, and identify critical nodes. Over time, this knowledge enables the adversary to strategically remove or compromise exactly those nodes that would most severely disrupt connectivity. Our results already demonstrate that targeted removal of just three key nodes significantly reduces network density and increases path length. 

The observed centralization score supports the percolation results, indicating that the network is becoming increasingly centralized, and the removal of key nodes may lead to a decrease in network density and a substantial increase in average path length.

\section{Conclusion}

In summary, by employing social network analysis, this study investigated the resilience of the I2P network to adversarial attacks. The findings reveal a vulnerability to attacks targeting central nodes, as their removal can disrupt network connectivity. The most influential nodes, characterized by a high number of connections, represent a critical point of failure. With the removal of the three most central nodes, we found that the average path length of the network increases while the density decreases. These findings suggest the I2P networks may be vulnerable to centrality-based attacks, offering valuable insights for future security enhancements.

\section{Limitation and Future Work}

The main limitation of this study is the dataset size. An attacker with fewer resources would still be able to observe more than the nodes included in this study. This introduces a sampling bias, as the partial view of the network may not fully capture the true structural complexity and resilience characteristics of the live I2P network. Consequently, the results may slightly overestimate or underestimate the network’s vulnerability under real-world adversarial conditions. In future work, a more robust dataset collected from the live I2P network should be utilized to validate the findings presented in this study.


\bibliographystyle{spmpsci} 
\bibliography{refs.bib} 

\begin{thebibliography}{10}
\providecommand{\url}[1]{{#1}}
\providecommand{\urlprefix}{URL }
\expandafter\ifx\csname urlstyle\endcsname\relax
  \providecommand{\doi}[1]{DOI~\discretionary{}{}{}#1}\else
  \providecommand{\doi}{DOI~\discretionary{}{}{}\begingroup \urlstyle{rm}\Url}\fi

\bibitem{i2pmetrics}
I2p metrics.
\newblock \urlprefix\url{http://i2p-metrics.np-tokumei.net/}.
\newblock Accessed: 2025-04-29

\bibitem{Abdo/978-3-031-53472-0_30}
Abdo, J.B., Hossain, L.: Modeling the invisible internet.
\newblock In: Complex Networks {\&} Their Applications XII, pp. 359--370. Springer Nature Switzerland, Cham (2024)

\bibitem{Almeira_centralized_attack}
Almeira, N., Billoni, O.V., Perotti, J.I.: Scaling of percolation transitions on erd\"os-r\'enyi networks under centrality-based attacks.
\newblock Phys. Rev. E \textbf{101}, 012,306 (2020).
\newblock \doi{10.1103/PhysRevE.101.012306}.
\newblock \urlprefix\url{https://link.aps.org/doi/10.1103/PhysRevE.101.012306}

\bibitem{Borgatti2005}
Borgatti, S.P.: Centrality and network flow.
\newblock Social Networks \textbf{27}(1), 55--71 (2005).
\newblock \doi{https://doi.org/10.1016/j.socnet.2004.11.008}.
\newblock \urlprefix\url{https://www.sciencedirect.com/science/article/pii/S0378873304000693}

\bibitem{BORGATTI2006}
Borgatti, S.P., Carley, K.M., Krackhardt, D.: On the robustness of centrality measures under conditions of imperfect data.
\newblock Social Networks \textbf{28}(2), 124--136 (2006).
\newblock \doi{https://doi.org/10.1016/j.socnet.2005.05.001}.
\newblock \urlprefix\url{https://www.sciencedirect.com/science/article/pii/S0378873305000353}

\bibitem{igraph_Csardi_2006}
Csardi, G., Nepusz, T.: The igraph software package for complex network research.
\newblock InterJournal \textbf{Complex Systems}, 1695 (2006).
\newblock \urlprefix\url{https://igraph.org}

\bibitem{Evans2022}
Evans, T.S., Chen, B.: Linking the network centrality measures closeness and degree.
\newblock Communications Physics \textbf{5}(1), 172 (2022).
\newblock \doi{10.1038/s42005-022-00949-5}.
\newblock \urlprefix\url{https://doi.org/10.1038/s42005-022-00949-5}

\bibitem{Hoang2018}
Hoang, N.P., Kintis, P., Antonakakis, M., Polychronakis, M.: An empirical study of the i2p anonymity network and its censorship resistance.
\newblock ACM (2018-10-31)

\bibitem{Hu}
Hu, Y., Zou, F., Li, L., Yi, P.: Traffic classification of user behaviors in tor, i2p, zeronet, freenet.
\newblock In: 2020 IEEE 19th International Conference on Trust, Security and Privacy in Computing and Communications (TrustCom), pp. 418--424 (2020).
\newblock \doi{10.1109/TrustCom50675.2020.00064}

\bibitem{LinkenLiu2019}
Liu, L., Zhang, H., Shi, J., Yu, X., Xu, H.: I2p anonymous communication network measurement and analysis.
\newblock p. 105–115 (2019).
\newblock \urlprefix\url{https://www.scopus.com/inward/record.uri?eid=2-s2.0-85076128384&doi=10.1007%2f978-3-030-34139-8_11&partnerID=40&md5=6b25cbedc399f8d6214c7a09a4c91703}

\bibitem{PeipingLiu}
Liu, P., Wang, L., Tan, Q., Li, Q., Wang, X., Shi, J.: Empirical measurement and analysis of i2p routers.
\newblock Journal of Networks \textbf{9}(9) (2014)

\bibitem{Roberto_Magan-Carrion}
Magán-Carrión, R., Abellán-Galera, A., Maciá-Fernández, G., García-Teodoro, P.: Unveiling the i2p web structure: A connectivity analysis.
\newblock Computer Networks \textbf{194} (2021).
\newblock \urlprefix\url{https://www.scopus.com/inward/record.uri?eid=2-s2.0-85105448281&doi=10.1016%2fj.comnet.2021.108158&partnerID=40&md5=8dc61d6ebf89233bcd2d0f6fe8ee85f6}

\bibitem{Siddique}
Muntaka, S.A., Abdo, J.B., Akanbi, K., Oluwadare, S., Hussein, F., Konyo, O., Asante, M.: Mapping the invisible internet: Framework and dataset.
\newblock arXiv preprint arXiv:2506.18159  (2025)

\bibitem{Siddique_resilience}
Muntaka, S.A., Bou~Abdo, J.: Resilience of the invisible internet project: A computational analysis.
\newblock Internet Technology Letters \textbf{8}(5), e70,119 (2025).
\newblock \doi{https://doi.org/10.1002/itl2.70119}.
\newblock \urlprefix\url{https://onlinelibrary.wiley.com/doi/abs/10.1002/itl2.70119}

\bibitem{Newman}
Newman, M.E.J.: Modularity and community structure in networks.
\newblock Proceedings of the National Academy of Sciences \textbf{103}(23), 8577--8582 (2006).
\newblock \doi{10.1073/pnas.0601602103}.
\newblock \urlprefix\url{https://www.pnas.org/doi/abs/10.1073/pnas.0601602103}

\bibitem{PENG201817}
Peng, S., Zhou, Y., Cao, L., Yu, S., Niu, J., Jia, W.: Influence analysis in social networks: A survey.
\newblock Journal of Network and Computer Applications \textbf{106}, 17--32 (2018).
\newblock \doi{https://doi.org/10.1016/j.jnca.2018.01.005}.
\newblock \urlprefix\url{https://www.sciencedirect.com/science/article/pii/S1084804518300195}

\bibitem{Rapisardi2022-homeostatic}
Rapisardi, G., Kryven, I., Arenas, A.: Percolation in networks with local homeostatic plasticity.
\newblock Nature Communications \textbf{13}(1), 122 (2022).
\newblock \doi{10.1038/s41467-021-27736-0}.
\newblock \urlprefix\url{https://doi.org/10.1038/s41467-021-27736-0}

\bibitem{Rstudio}
{RStudio Team}: RStudio: Integrated Development Environment for R.
\newblock RStudio, PBC., Boston, MA (2020).
\newblock \urlprefix\url{http://www.rstudio.com/}

\bibitem{Seshadhri_Transitivity}
Seshadhri, C., Pinar, A., Durak, N., Kolda, T.G.: Directed closure measures for networks with reciprocity.
\newblock Journal of Complex Networks \textbf{5}(1), 32--47 (2016).
\newblock \doi{10.1093/comnet/cnv032}.
\newblock \urlprefix\url{https://doi.org/10.1093/comnet/cnv032}

\bibitem{JP2012}
Timpanaro, J.P., Chrisment, I., Festor, O.: A bird's eye view on the i2p anonymous file-sharing environment.
\newblock p. 135–148 (2012).
\newblock \urlprefix\url{https://www.scopus.com/inward/record.uri?eid=2-s2.0-84871531288&doi=10.1007%2f978-3-642-34601-9_11&partnerID=40&md5=934f8c0bab63b44084a9e0eb7b231ae9}

\bibitem{Wang}
Wang, X.B., Liu, P.P., Li, C.L., Tan, Q.F.: Towards measurement on the i2p network.
\newblock p. 2223–2228 (2013).
\newblock \urlprefix\url{https://www.scopus.com/inward/record.uri?eid=2-s2.0-84886376293&doi=10.4028%2fwww.scientific.net%2fAMM.427-429.2223&partnerID=40&md5=970703aea504f2e13ef796378aada582}

\bibitem{wilson2016forensic}
Wilson, M., Bazli, B.: Forensic analysis of i2p activities.
\newblock In: 2016 22nd International Conference on Automation and Computing (ICAC), pp. 529--534. IEEE (2016)

\end{thebibliography}
\end{document}